\documentclass{aa}
\usepackage{graphicx}

\begin{document}

\title{An efficient parallel tree-code for the simulation\\
of self-gravitating systems\thanks
{Supported by CINECA (http://www.cineca.it) and CNAA (http://cnaa.cineca.it)
under Grant \emph{cnarm12a\/}.}}

\author{P. Miocchi \and R. Capuzzo-Dolcetta}
\institute{Dipartimento di Fisica, Universit\'a di Roma ``La Sapienza",
(Ed. Fermi) P.le Aldo Moro, 5, I00185 -- Rome, Italy.}

\offprints{P. Miocchi,
\email{miocchi@uniroma1.it}}

\date{Received ??? / Accepted ???}

\abstract{
We describe a parallel version of our tree-code for the simulation of
self-gravitating systems in Astrophysics.
It is based on a dynamic and adaptive method for the
domain decomposition, which exploits
the hierarchical data arrangement used by the tree-code.
It shows low computational costs for the parallelization overhead
 -- less than 4\% of the total CPU-time in the tests done --  because the domain
decomposition is performed `on the fly' during the tree setting and
the portion of the tree that is local to each processor `enriches'
itself of remote data only when they are actually needed.
The performances of an implementation of the parallel code on a Cray T3E
are presented and discussed.
They exhibit a very good behaviour of the speedup ($=15$ with $16$ processors
and $10^5$ particles) and a rather low load unbalancing ($< 10$\% using up to
16 processors), achieving a high computation speed in the forces evaluation
($>10^4$ particles/sec with 8 processors).
}

\titlerunning{An efficient parallel tree-code}
\maketitle

\section{Introduction}
Tree-codes (Barnes \& Hut \cite{BH}, hereafter BH; Hernquist \cite{H}) are particle
algorithms extensively employed
in the simulations of large self-gravitating astrophysical
systems. They have the capability to speed up the numerical evaluation of the
gravitational interactions among the $N$ bodies of the system.
Of course, the parallelization of the algorithm aims at attaining larger and larger
$N$ in the numerical representation of the real system; this is
important not only to improve the spatial resolution, but also to get much more
meaningful results, because a too low number of `virtual' particles in comparison with
the number of real bodies, gives rise to a unphysical shortening of the 2-body
collisions time.

In general, an efficient parallelization means a data distribution to the
processors, the so called \emph{domain decomposition\/} (DD), so as to i)
distribute the numerical work as uniformly as possible, ii) minimize the
data exchange among the processors (hereafter PEs). Of course, this latter point is
relevant only on distributed memory platform. Moreover, such DD should be performed
with a minimal computational cost.

In the numerical evaluation of the gravitational interactions it is difficult to
deal with these tasks, because the long-range nature of gravity makes data
transfer among PEs unavoidable.
Furthermore, self-gravitating systems have often non-uniform mass
distributions, that give rise to very inhomogeneous distributions of the
work-load (the amount of calculations needed to evaluate the acceleration of a
particle).
This implies that the DD should be weighted, in some way, according to the work-load.

Finally, the hierarchical arrangement of the subsets of the mass distribution
which the tree-code is based on, implies that most of the computations for
evaluating the acceleration of a particle regard the evaluation of the force due to
\emph{close\/} bodies. This suggests a \emph{spatial\/} DD:
each domain should be enclosed in a volume as \emph{contiguous\/} and compact as
possible.

At present, one of the most used approach for the DD is the orthogonal recursive
bisection (Warren \& Salmon \cite{orb}; Dubinski \cite{dubinsky}; Lia \& Carraro
\cite{carraro}; Springel et al. \cite{gadget}),
which consists in a recursive subdivision of the space in pairs of sub-domains
equally weighted.
On every sub-domain, the owning PE builds (independently of the others) its
\emph{local\/} tree data structure. Such structure is then enlarged enclosing those
data, belonging to \emph{remote\/} trees, that are needed to the local forces
evaluation.

The main disadvantage of this approach is that the retrieval of remote data
is complicated (and computationally expensive, too) mainly because
of the lack of an addressing reference scheme common to all the PEs.
The `hashed oct-tree' method (Warren \& Salmon \cite{hashed}, hereafter WS)
solves this
problem, but a certain implementation complexity still remains and some radical
changes are required in the way the tree arrangement is usually stored.
For this reason we decided to carry out a new and easy to implement
method for sharing efficiently the computational domain among PEs
in a distributed memory architecture.

This paper is organized as follows.
In Sect. \ref{treecode} we describe the differences of our tree-code in respect with
the original method illustrated in BH,
both from the general point of view of the
algorithm and in connection with the parallelization approach.
This latter is described in Sect. \ref{parallelization} for both the stages
the tree-code is made up of. Finally, the performances of a PGHPF
(Portland Group High Performances Fortran) implementation running on a
Cray T3E computer are discussed in Sect. \ref{results}.
\section{Our version of the BH tree-code}
\label{treecode}
In this Section we describe some modifications of the original BH tree-code,
which are also important
(as we will see later)
from the point of view of the parallelization technique.
They regard the construction of the tree arrangement.
The reader who is not familiar with the basical features of
tree-codes, can found detailed descriptions in Warren \& Salmon (\cite{orb});
Hernquist (\cite{H}); Hernquist \& Katz (\cite{HK}); Springel et al.
(\cite{gadget}).

Let us give some definitions that may differ from those used by
other authors:
i) the \emph{boxes\/} are the cubes that make up the
hierarchical structure (arranged as an octal tree graph) built during the
\emph{tree-setting\/} stage by subdividing recursively the \emph{root\/} box
enclosing all the system, ii) the root is at the $0$-th \emph{subdivision level\/}
and iii) a \emph{parent\/} box, at the $l$-th level, is a box which includes more
than one particle and which is splitted into 8 cubic \emph{sub-boxes\/} at the
$(l+1)$-th subdivision level,
iv) the \emph{terminal\/} boxes are those with just one particle inside and v)
the \emph{tree-traversal\/} is the phase in which all the particles accelerations
are evaluated by ``ascending'' the tree from the root upward.

We adopted an internal memory representation of the tree structure
that makes use of pointers, i.e. integers pointing to the locations in
which the data of the sub-boxes have been stored.
This allows to accede recursively to boxes' data with a $O(\log N)$ order of
operations and, moreover, it permits, as we will see, to complete easily and
with a minimal communications overhead the portion of the tree initially
assigned to a given PE, appending those remote box data needed to calculate
the accelerations on the particles in its domain.

The tree-setting is performed through a recursive method
different from the original approach used in BH.
The method can be outlined as:
given a \emph{parent\/} box and the set of all the particles it contains,
the subset of the particles enclosed in a given sub-box is found.
If it is \emph{non}-empty then the multipolar coefficients of the sub-box, plus
various parameters, are evaluated and stored into a free memory location.
Then a pointer in the parent box is set to point to such location.
This procedure starts from the root box and is repeated recursively for any
non-terminal sub-box.
Also for the evaluation of the multipolar coefficients the
recursive `natural' approach is used (as in Hernquist \cite{H}).

Within the framework of the tree-setting just described, it is important
to employ a fast method to check whether a particle belongs to a box or not.
In this respect, we implemented a \emph{spatial mapping\/} of the particles, which
`translates' the coordinates of each of them into one binary number
(a `key'), enclosing all the
necessary informations about which box contains it at \emph{any\/} subdivision
level. Moreover, it is quite easy to get quickly such
informations using binary operations within a \emph{recursive\/} context.
Details about such method are given in Appendix~\ref{mapping}.
\section{The parallelization method}
\label{parallelization}
\subsection{Parallel tree-setting and domain decomposition}
\label{par_tree_setting}
As far as the parallel execution of the tree-setting is concerned, an important
feature of the logical data structure is that the lower levels of the tree
are made up of few but highly populated boxes while, on the contrary, at upper
levels there are many boxes, but containing few particles.

\begin{figure*}
\centering
\includegraphics[width=12cm]{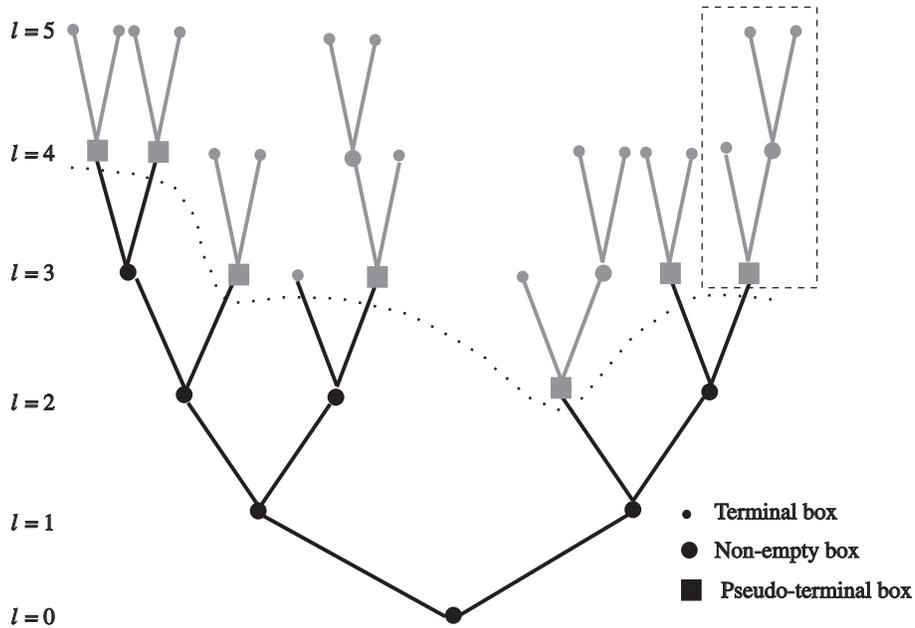}
    \caption{Example of types of boxes for a binary tree. Upper (gray)
and lower (black) parts of the tree are depicted (up to the 5$^{th}$ level).
The dashed rectangle encloses one of the `sub-trees' belonging to a processor's
domain.}
\label{treeparts}
\end{figure*}

This suggests two different schemes of work distribution to the PEs during
this phase.
Indeed, in order to have a good load-balancing, it is desirable to make the
number of the `computational elements' much larger than the
number $p$ of the PEs which the work has to be distributed to.
Hence, in our case it is convenient that, on one side the parallel setting up
of the lower levels of the tree is done by assigning to each processor a
sub-set of the particles belonging to the \emph{same\/} box; on the other side,
for the construction of the upper levels, whole sets of particles belonging
to \emph{distinct\/} boxes should be assigned to each PE.
Thus, before going any further, let us give some useful definitions.
Given $k > 1$ a fixed integer, we call
\begin{itemize}
\item \emph{lower\/} box: a box containing a number of particles $n$ such that $n > kp$;
\item \emph{upper\/} box: a box with $n \leq kp$;
\item \emph{pseudo-terminal\/} (PTERM) box: an upper box having a parent lower
box.
\end{itemize}
Finally, we call `lower (upper) tree' the portion of the whole tree
made up of lower (upper) boxes (see Fig.~\ref{treeparts}).
\noindent

The parallel tree-setting is then executed in two steps, the first step for the
construction of the lower tree and the second one for that of the upper tree,
as described in the following Sections.
\subsubsection{Lower tree setting}
\label{low-tree-setting}
In the first step, the particles are initially distributed at random to the PEs,
i.e. without any correlation with their spatial location.
Then, the PEs start building the tree, working on lower boxes, according to the
recursive procedure described in Sect. \ref{treecode}. They consider, at
the same time, the \emph{same\/} box but dealing, of course, only with their own
particles.
This phase of the tree setting stops at PTERM boxes (instead of at terminal ones,
as it is done in the serial version) where no more `branches' are set up.
In order to obtain an efficient parallel execution, the evaluation
of the multipolar coefficients of (lower) boxes is done directly, i.e.
by means of summations running over the set of particles contained, rather than
by the recursive formulas involving the sub-boxes coefficients.

To attain the maximum data-locality, each PE keeps a copy of the tree structure
in its own local memory.
This makes that, in the tree-traversal stage, the reading access to
lower boxes\footnote{They are very frequently met in evaluating
particles' accelerations, because they generate the long-range field.} data
will not be slowed down continuously by the inter-processors data transfer, which is
one of the performances bottlenecks of codes running on distributed memory parallel
computers.
Note also that the amount of local storage needed to this part of the tree scales
like the number of lower boxes, that is like $\sim \tau\log \tau$, being $\tau \sim
N(kp)^{-1}$ the total number of PTERM boxes.
Thus, the memory occupation for each local copy
of the lower tree scales, conveniently, like the number of particles \emph{per\/}
processor.
During the current step, the large number of particles in lower boxes
(provided that $k$ is sufficiently great), together with the random
particle distribution to the PEs, ensures a good work-load balancing.

\noindent
\begin{figure}
 \centering
\includegraphics[width=6cm]{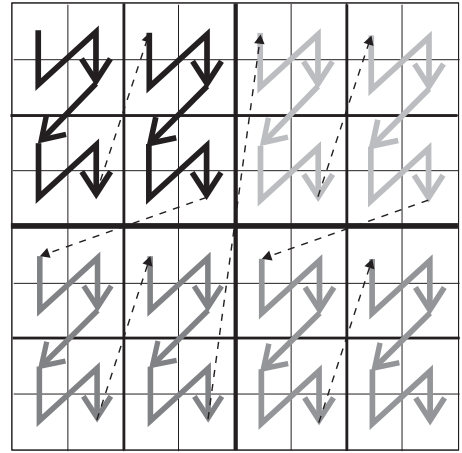}

 \caption{Example of PTERM boxes inspection order, for a uniform particle
distribution in 2-D.
The boxes are on the 3$^{\rm rd}$ level of the spatial subdivision
and each pattern corresponds to a different processor's domain (among 4 PEs),
while the dashed arrows indicate the `jumps' along the path.}
\label{order}
\end{figure}

At this point a suitable re-distribution of the particles (for an efficient
tree-traversal) is performed `on the fly' by exploiting our recursive way to
set up the logical tree structure.
Actually, such recursive approach, together with the technique used in mapping
the particles' coordinates (see Fig.~\ref{sub-box}),
leads to a particular order in which the PTERM boxes are met.
Such order corresponds to a one-dimensional path connecting a PTERM box to another
spatially \emph{adjacent}\footnote
{Actually, along the path there are some discontinuities in form of `jumps' from a
PTERM box to another non-adjacent one, which could be avoided with a more
complicated (non self-similar) order, as described in WS.}
in a self-similar fashion (see Fig.~\ref{order}). Though similar to that used by
WS, the order is obtained in a substantially different way, as discussed in Appendix
\ref{mapping}.

This order is such that, by cutting the
path into $p$ contiguous pieces with the same `length' and then assigning to the
$i$-th PE the particles contained into all the PTERM boxes in the $i$-th piece,
one obtains a particularly efficient DD. Indeed, it is characterised by
having most of sub-domains compact in space
\footnote{The efficiency comes from that, in evaluating the acceleration on a
particle, most of interactions are with the closest bodies. In fact, the
number density of the bodies satisfying the opening
criterion at a distance $d$ from the particle is roughly $\propto
(\theta d)^{-3}$ (with $\theta$ the open-angle parameter)
which decreases very rapidly with the distance.
Hence a compact sub-domain will contain most of such bodies.}.

The particles assignement is performed every time a PTERM box is met.
Moreover, other than the particles contained, also the data regarding
the box are stored into the \emph{local\/}
memory of the PE. For this reason,
it is necessary that the pointer in the parent box pointing
to the PTERM one, includes also the information
about which PE owns this latter. This way, any other PE can easily accede to
the box data.
The information is included within the pointer itself by constructing the
\emph{full-address\/} of the PTERM box, as
described in Appendix~\ref{fulladdressing}.

As in WS, we found that a good load-balancing can be
attained if one
`measures' the path length in terms of the computational loads of the PTERM boxes,
defining such quantity as the sum of the `weights' of all the particles
inside them, being the particle weight proportional to the number of
bodies (both particles and boxes) whose force on it has been evaluated during the
tree-traversal of the \emph{previous\/} time step.

\begin{figure}
\centering
\includegraphics[width=7cm]{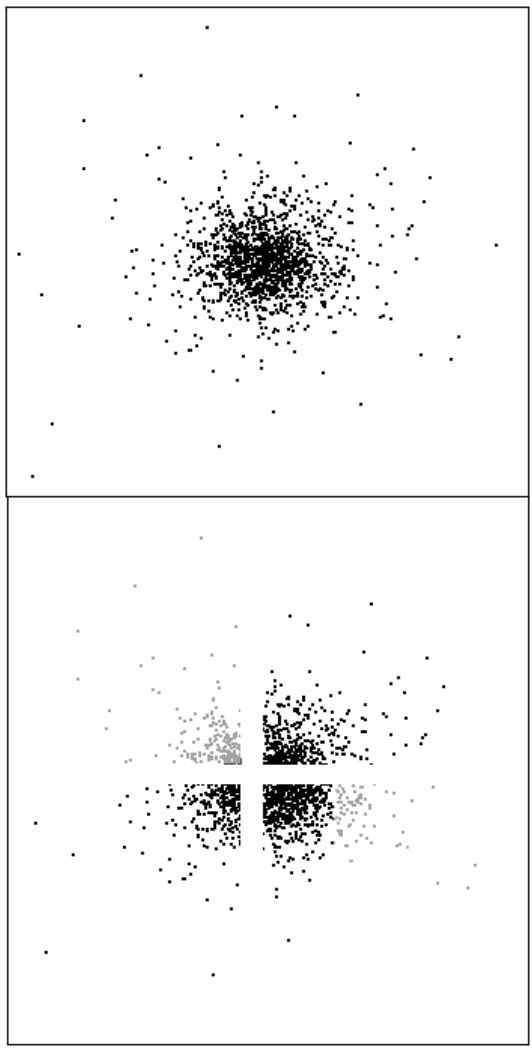}
    \caption{
Example of domain decomposition among four PEs, for a cluster represented
with 16,384 particles. Top: `section' of the 3-D particles distribution
lying on the $yz$ plane. Bottom: the sections of the 4 sub-domains
have been spaced for 
clarity; note how one of them (the grey one) is not completely compact.}
\label{domains}
\end{figure}

An important difference with respect to the WS' method is that in our scheme the DD
is performed via a (PTERM) \emph{boxes\/} distribution to the PEs, rather then by
directly distributing
particles, and this means an easier parallel tree construction of the sub-trees
in comparison with the HOT method. In this latter, in fact, there are several
complications in setting up the local parts of the tree, such as: the `broadcasting'
of branch boxes, the inter-processor exchange of data regarding particles sited
at the border of sub-domains, etc.
In Fig.~\ref{domains} an example of particles distribution to 4 PEs is plotted for a
non-uniform case.

\noindent
\subsubsection{Upper tree-setting}
The second step is the construction of the remaining
upper part of the tree, which is performed according to the same recursive
procedure used for lower boxes
but, in this case, every PE works independently and without synchronism, starting
every time from each PTERM box in its own domain.
Every PE will store, in its local memory, the logical and data structure of all the
sub-trees whose roots are given by such boxes.
For example, in Fig.~\ref{treeparts}, the PE owning the rightmost PTERM box, will
set up the sub-tree enclosed within the dashed rectangle. Another difference,
in comparison with the lower tree setting, is that all the pointers box
$\rightarrow$ sub-box adopt the full-addressing, because, in principle, any box
could be required by other processors during the tree-traversal.


\subsection{Parallel tree-traversal}
\label{par_tree_trav}
In this stage, each PE uses exactly the same recursive procedure usually
adopted in serial tree-codes (see e.g. Hernquist \cite{H}) to evaluate the
forces acting on the particles though, of course,
only on those belonging to its domain.

At the beginning of this phase, such domain includes
both the whole lower tree and those sub-trees whose roots are the PTERM boxes
assigned to the PE.
We can call all this set of boxes the initial \emph{locally essential tree\/} (LET).
This latter is not yet ``complete'', in the sense that it does not include,
yet, all the bodies that are necessary to evaluate the forces on the particles
in the PE's domain,
lacking some of the upper boxes belonging to \emph{remote\/} sub-trees
(though they are the minority of all the required bodies, thanks to
the spatially compact DD).

Anyway, the suitable addressing scheme adopted and the belonging of the boxes to the
\emph{overall\/} tree topology (as well as the recursive approach for the
tree-traversal), allow us to perform the \emph{LET completion\/} `at run
time'.
Given a particle belonging to a certain PE and given a box
$B\in$ LET whose sub-boxes have to be handled: if a sub-box does not belong
to the LET, as can be immediately recognized by Eq.~\ref{fulladdr2}, then
(i) get all its data (and all its pointers) from the owning PE's memory at
the address given by Eq.~\ref{fulladdr3},
(ii) copy them into a free location of the local memory, (iii) change the
pointer in $B$ to point this new \emph{local\/} address.
This way, the sub-box is included into the LET and when any other particle
requires it, it is already found into the local memory.
This mechanism minimizes the amount of inter-processor communications.

Note that the full-addressing mechanism makes remote data
retrieval immediate, thanks to that the local \emph{sub}-trees are portions
of a \emph{single\/} and global tree arrangement, unlike the orthogonal recursive
bisection scheme (Warren \& Salmon \cite{orb}).
In our opinion our addressing method is as `global' as that used in WS,
though easier to be implemented and with a lower computational overhead,
as we will see.

\section{Results}
\label{results}
The efficiency of the parallelization method has been checked by
analysing the performances for a \emph{single\/} evaluation
of the forces on a set of particles. Such evaluation includes one tree-setting
(including DD) and one tree-traversal step, as well as all the necessary
inter-processor communications and remote accesses (LET completion),
\emph{without\/} the time integration of trajectories.
Indeed, it is known that most of the CPU-time used by a simulation of large
self-gravitating systems is spent in the computation of the
gravitational interactions. Usually this latter takes at least
80\% of the total CPU-time, while the time advancing of particles' dynamical
quantities (position, velocity, etc.) takes just $\sim$ 10\%, because
its computational cost scales like $N$. This cost is even smaller for
low-order time integration schemes, like those generally used in conjunction
with tree-codes\footnote{
The commonly accepted degree of accuracy in the force calculation is about
one part in $10^2$--$10^3$. This makes high-order time schemes superfluous (at
least in not too dynamic situations).}.
Moreover, it is generally very simple to parallelize time integration methods,
because the time advancing of a particle is independent of
that of the others and the corresponding work-load is, normally, very
homogeneous.
For this reason our tests involve the forces computation
only, which indeed represents the key problem to overcome for getting a good
parallelization.

Nevertheless, one has to be careful when time integration algorithms adopt
individual time steps (that is a desirable feature when dealing with
self-gravitating systems with a wide range of time scales), because
they imply a force evaluation which is mostly performed on a \emph{subset\/} of
the entire set of particles. We discuss this problem in Appendix~\ref
{time_integration}.

\begin{figure}
\centering
\includegraphics[width=7cm]{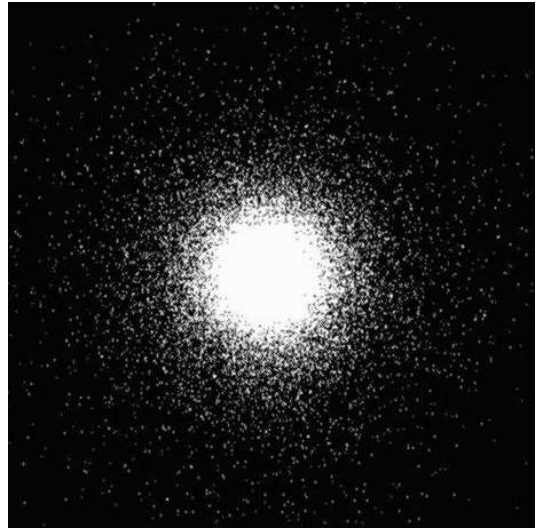}
    \caption{
The clustered set of 128K particles used in the tests.}
\label{plummer}
\end{figure}

All the tests were performed on a set of $N=128$K equal mass ($m$) particles
distributed according to the Plummer profile (known to fit acceptably globular
clusters, at least in regions not too far from the center, see Binney \&
Tremaine \cite{binney})
$\rho(r)=\rho_0(1+r^2/r_c^2)^{-5/2}$, within a sphere of radius $R$ such to
include a particles total mass, $M=Nm$, such that $M=0.995\times M_\infty$,
where $M_\infty=\int_0^\infty 4\pi r^2\rho dr$.
The core radius is chosen as $r_c=6\times 10^{-2}R$, while
$\rho_0=3M_\infty(4\pi r_c^3)^{-1}$ is the central
density. This is a highly non-uniform distribution,
being $\rho_0/\rho(R)\sim 10^6$ (see Fig.~\ref{plummer}).

For the box-particle force evaluation we used multipolar series truncated at
the quadrupoles, and the original BH `opening' criterion
with different values for the open-angle parameter $\theta$.
The maximum number of particle in PTERM boxes
(see Sect. \ref{par_tree_setting}) was set equal to $16\times p$,
this value giving the best performances for any number of PEs ($p$), as we
checked.
The tests ran on a Cray T3E and regarded our parallelization method implemented
by means of PGHPF/Craft directives.
\noindent
\subsection{The performances}
The good efficiency of the parallelization approach described in previous
Sections is basically shown by three facts: i) a behavior of the relative
speedup\footnote{The rapidity gain one has when going from a run with one PE to
the same run with $p$ PEs.} close to the ideal one (linear in $p$);
ii) a low unbalancing of the work-load; iii) a low parallelization
overhead\footnote{The CPU-time needed by all those instructions which would not
be necessary in a \emph{serial\/} execution.}.

\noindent
\begin{figure*}
\centering
\includegraphics[width=11cm]{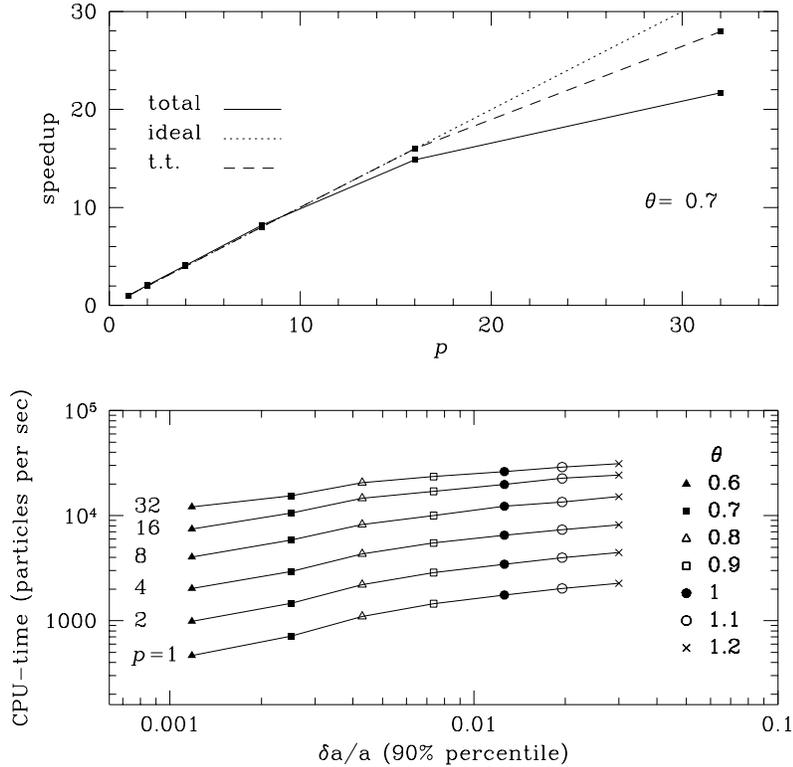}
    \caption{
Speedup for the force evaluation on $N=128K$ particles (Plummer profile).
Bottom: code speed vs. the error on the forces $\delta a/a$ at the 90\% percentile
(i.e. such that 90\% of particles have a force affected by a relative error
$\leq \delta a /a$) for various number of PEs ($p$) and different values of
$\theta$, as labeled.
Top: the speedup with respect to a single PE run (in the $\theta=0.7$ case)
plotted for the overall force evaluation and for the tree-traversal (t.t.)
only.}
\label{speedup}
\end{figure*}
The good overall code scalability is shown in the upper panel of
Fig.~\ref{speedup}.
It appears to be rather good for $p\leq 16$ (i.e. $N/p\geq 8192$).
For more than 16 PEs, the performances start to degrade because of the too small
number of particles \emph{per\/} PE: with $p\leq 32$ one has $\leq 4096$ particles
per PE that makes the tree-code not that efficient in itself.
To give immediate indications of how fast the calculations are for a given accuracy,
we show in the lower panel the absolute speed of the code
versus the relative error on the forces evaluation.
The relative error on the evaluation of the force (per unit mass) on a particle
 -- due to the use of the truncated multipolar expansion for the box satisfying the
opening criterion --  is defined as:
$\delta a/a \equiv |a_{\rm tc}-a|/a$, where $a$ is the magnitude
of the acceleration evaluated by means of the `exact' particle-particle summation,
and $a_{\rm tc}$ denotes the magnitude of the acceleration calculated
via the tree-code.

\begin{figure}
\centering
\includegraphics[width=8cm]{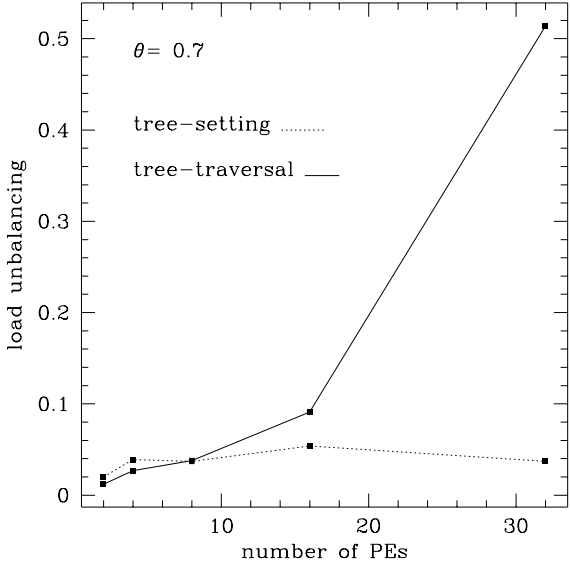}
    \caption{
Load unbalancing parameter ($u$) for a single force evaluation with $\theta=0.7$.
Solid line: for the tree-setting stage; dotted line:
for the tree-traversal.}
\label{workload}
\end{figure}

The computational work-load is well balanced among the PEs.
A natural way to quantify the load unbalancing, $u$,
is via the formula $u=(t_{\rm max}- t_{\rm min})/<\!\! t\!\! >$,
where $t_{\rm max}$ and $t_{\rm min}$ are, respectively, the maximum and the minimum
CPU-time spent by the PEs to perform a given procedure and $<\!\! t\!\! >$ is the
averaged CPU-time.
From Fig.~\ref{workload}, we can see that, for a sufficiently high number of
particles per PE (say for $N/p \geq 8,000$) we have a quite low $u$,
that is less than 10\% for the tree-setting and always less than 6\% during
the tree-traversal, demonstrating the efficiency of the DD. Only for $N/p\sim 4,000$
the unbalancing becomes unacceptable ($>50$\%).
\noindent

Last, but not least, we can see in Table \ref{tabella} that the
parallelization overhead
takes only 3.2\% of the total CPU-time in a 8 PEs run.
Moreover, as we verified, this percentage increases significantly only when
$N/p< 8,000$.
Thus, in optimal conditions the `surplus' of CPU-time specifically
needed to make the parallelization operative
(in our case spent by the DD plus the LET completion) is almost negligible.
This point is crucial in order to state that a parallel code is really
efficient.

Indeed, even in a distributed memory context one can get a
highly scalable tree-code with a good work-load balancing, using
just an absolutely banal DD.
In fact, the load balancing can be achieved by means of a `dynamical'
distribution\footnote{All the particles to be processed are put in a `queue'.
As soon as a PE has finished its previous work, it gets the first particle
in the queue and evaluates its acceleration.} of
the particles to the PEs (see Singh et al. \cite{singh}), tolerating a great deal of
communications and remote accesses.
Following this approach, we implemented another parallel version of the
tree-code obtaining a good speedup scaling and a low load unbalancing on the T3E,
but the communications overhead \emph{heavily\/} affected the absolute performaces
making it not convenient for practical use (see Capuzzo-Dolcetta \& Miocchi
\cite{cdm-ssc97}, \cite{cdm-cpc}).
\begin{table}[ht]
\caption{Code CPU-time consumption with 8 PEs, 128K particles and
$\theta=0.7$. Italic: parallelization overhead.
\label{tabella}}
\begin{tabular}{lrr}
\hline
Code section&sec&\% \\
\hline
tree-setting&$1.4$&$6$\\
\ \ \emph{domain decomposition}&$0.1$&$0.5$\\
tree-traversal&$21$&$94$\\
\ \ LOWER tree-traversal&$3.3$&$15$\\
\ \ UPPER tree-traversal&$18$&80  \\
\ \ \ \ 
\emph{LET completion}&$0.6$&$2.7$\\
total&$22.4$&$100$\\
\hline
\end{tabular}
\end{table}

Finally, the code memory usage requires roughly 1 Kbyte per particle. For
instance, more than $10^7$ particles can be handled by 128 processors
having 128 Mbyte each. Such amount of particles can be furtherly increased
in a more optimized message passing implementation.
\subsection{Comparisons with other codes}
To make really significant comparisons of the performances of different
tree-codes, one should ensure the forces computation be done with the same accuracy
and on the same set of particles. Of course, such performances depend also on the
opening criterion adopted, because at a given accuracy and with a given set of
particles, different opening criteria can give different amounts of interactions to
evaluate on a particle, thus giving different computation speeds.
Therefore, if one wants to compare specifically the efficiency of the
\emph{parallelization\/} approach, then the tests should be done with the same opening
criterion too\footnote{In general the implementation of the opening criterion
does not really affect the parallelization method, and it is
a rather simple task as well.}.

Unfortunately, it is often very difficult to make such conditions hold
with the tree-codes available in the literature. For this reason we
decided to compare codes speed at a given amount of \emph{computational work\/} done
to evaluate forces. This makes the comparison independent of: the particles
distribution, the number of particles, and the accuracy (i.e. the
opening criterion and its parameters).
In tree-codes the amount of numerical work done on a given particle, $w_i$, is
naturally quantified as the number of `interactions' evaluated to estimate the
force on it (as in the particle work-load definition of Sect. \ref{low-tree-setting}),
namely the total number of bodies (both boxes and single particles) of which
the tree-code evaluates the force they exert on the particle itself (in a
particle-particle method one would have $w_i=N-1$).

\begin{figure}
\centering
\includegraphics[width=8cm]{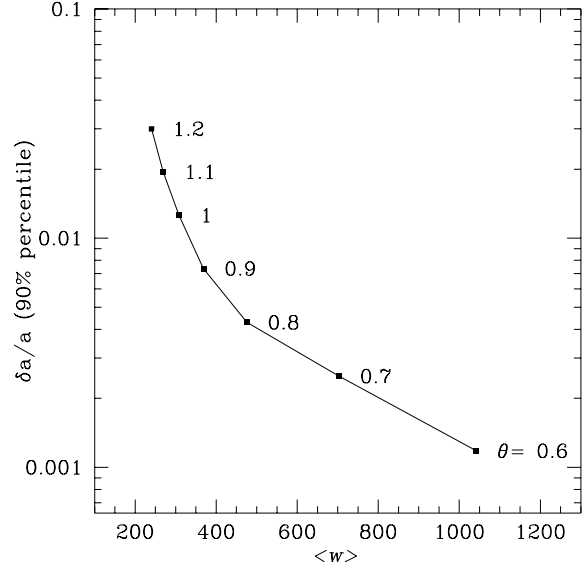}
\caption{
Relative error on forces evaluation (at 90\% percentile) versus the averaged
computational work, using the BH opening criterion. The corresponding values for
the open-angle parameter $\theta$ are labeled.}
\label{work}
\end{figure}

In Fig.~\ref{work} we plotted the error on the forces evaluation ($\delta a/a$),
versus the averaged computational work, $<\!\! w\!\!>=(\sum_i w_i)/N$, needed by
our code to evaluate the forces on the set of $N=128$K particles above-described.
This allows us to compare `honestly' our code performances
with those of other codes.
\noindent

In Springel et al. (\cite{gadget}) the authors tested their tree-code (GADGET) on a Cray
T3E. They give speed measurements for a rather clustered cosmological
distribution, using the BH opening criterion with $\theta=1$. In such
conditions their code gives $\delta a /a\sim 3.5\times 10^{-2}$ at 90\%
percentile, with $<\!\! w\!\!>\simeq 200$ interactions per particle.
From Fig.~\ref{work} we can see that, with the Plummer distribution we
used, the closest value for $<\!\! w\!\!>$ is achieved for
$\theta =1.2$, which gives $<\!\! w\!\!>\simeq 230$ interactions per
particle and corresponds to an accuracy of $\delta a /a\sim 3\times 10^{-2}$.
Such accuracy is obtained in a run that, using e.g. 8 PEs, is
performed with a speed of $15,000$ particles/sec (as shown in the lower
panel of Fig.~\ref{speedup}). Note that such run
includes, apart from the tree-traversal, also the tree-setting and all
the overhead needed to a parallel execution.
At the same conditions, but performing \emph{only\/} the tree-traversal
stage, GADGET has a lower speed: about $13,000$ particles/sec (with 8 PEs).
It is worth noting that it shows a
load unbalancing of about 19\% with $N/p\sim 3\times 10^4$, while our code
exhibits $u\sim 4$\% with the same ratio $N/p$.
One has to say, finally, that GADGET has been implemented using
message passing instructions, certainly a more suitable approach with
respect to that we adopted (see next Section).

Another comparison can be done with the parallel code illustrated in
Dubinski (\cite{dubinsky}). In this case the code ran on a Cray T3D and the author
employed a more efficient modified BH opening criterion (Barnes \cite{barnes}).
Anyway, the rapidity of the code is given both in terms of the total
computational work ($N<\!\! w\!\! >$) performed in one second, and in terms
of particles per second.
With 16 PEs such code evaluated about $3\times 10^6$ interactions/sec,
corresponding to $6,000$ particles/sec, for a forces computation performed on
a cluster with $N=1.1$M particles.
This means that the code spent $t\simeq 180$ sec in that run. Hence,
being $N<\!\! w\!\! >/t\simeq 3\times 10^6$, it handled $<\!\! w\!\! >\simeq 500$
interactions per particle.
With our tree-code such $<\!\! w\!\! >$ corresponds to an accuracy of about
$\delta a/a\sim 4\times 10^{-3}$ (Fig.~\ref{work}), and so it is performed at
$\sim 13,000$ particles/sec with 16 PEs (Fig.~\ref{speedup}).
Of course, our speedup factor of $2$ is partially due to the improved
performances of the T3E compared with the T3D, even if the direct message
passing approach used by the author is more efficient then the use of the
PGHPF compiler.


\subsection{Remarks about the implementation}
The use of PGHPF/Craft directives is certainly not the best
way to implement a parallel tree-code on a distributed memory
architecture, but we did so in order to obtain quickly a
ready-to-use version. Indeed, the directives permit just to
distribute in a \emph{simple\/} way loop iterations to the PEs, as well as the
elements of shared arrays to their local memory, without using \emph{explicit\/}
message passing routines.

The highest price to pay for such simplicity, is that the way message
passing operations are actually performed cannot be controlled and optimized.
In our specific case, each PE has to \emph{copy\/} all its local upper tree to
logically shared arrays, in order to enable the other PEs to accede to
it during the tree-traversal. This means a considerable waste of memory
(and communications), which can be avoided in a direct message passing
implementation, so to reduce furtherly the parallelization overhead.

Similar storage and clock time wastings are due also to that the PGHPF
compiler is generally not capable to recognize local references within a
shared array, which are handled as if they were remote instead.
Anyway, we experimented also a slow local referencing, due to a
non optimal cache-memory managing.
For these reasons, our next goal is the development of an MPI version,
which would also be easily implementable on different distributed memory
platforms.

Finally, we have also carried out an implementation suitable for running
on a \emph{shared\/} memory computer.
In this case, of course, the DD has to take into account only the
work-load balancing and a `dynamic' particles distribution can be adopted
during the tree-traversal. Anyway, it turns out to be worth subdividing
the tree into upper and lower boxes for a balanced tree-setting.
Such implementation\footnote{We are thankful to the CASPUR center (sited at
the Universit\'a di Roma ``La Sapienza'') for the resources provided.}
was carried out using OpenMP directives on a SUN Enterprise 4500 HPC machine,
with 14 PEs.
The results are very good and, given the same parameters, we verified a
$40$\% speedup in comparison with the T3E PGHPF implementation.
\appendix
\section{Particle mapping}
\label{mapping}
Particles' locations are mapped converting
each coordinate\footnote{With the origin at a root box' vertex
and the axes parallel to its edges.}, $x_i$, $i=1,2,3$, into an
integer triple $q_i=\lfloor x_i\times 2^{l_{\max}}/L\rfloor$, where
$\lfloor...\rfloor$ indicates the truncation to an integer,
$L$ is the root box' size and $l_{\max}$ is the maximum subdivision level
\emph{a priori\/} allowed.
Then $q_{1,2,3}$ are combined into an integer number (the `key') $Q\in
[0,8^{l_{\max}}-1]$, which is defined as:
\begin{eqnarray}
Q&=&\sum_{l=1}^{l_{\max}}
\left[\mbox{mod}(\lfloor q_1/k_l\rfloor,2)+2\times\mbox{mod}
(\lfloor q_2/k_l\rfloor,2)+ \right.\nonumber \\
&{}&\left.+4\times\mbox{mod}(\lfloor q_3/k_l\rfloor,2)\right]^{k_l^3}, \label{Q}
\end{eqnarray}
where $k_l=2^{l_{\max}-l}$.
Despite its complicated definition, $Q$ can be
rapidly evaluated by means of direct bit manipulation routines\footnote{
For instance, a multiplication of an integer, $n$, by $2^j$, with $j>0$ ($j<0$),
corresponds to shift its binary representation by $j$ positions left (right),
whereas $\mbox{mod}(n,2)\equiv$ the least significant (rightmost) bit.
Hence $\mbox{mod}(n/2^m,2)$, with $m\ge 0$, is the bit in position $m$,
being the rightmost one in position 0. In FORTRAN such bit is given by
{\tt ibits($n$,$m$,1)}}
available both in FORTRAN and in C.

\begin{figure}
\resizebox{\hsize}{!}{\includegraphics{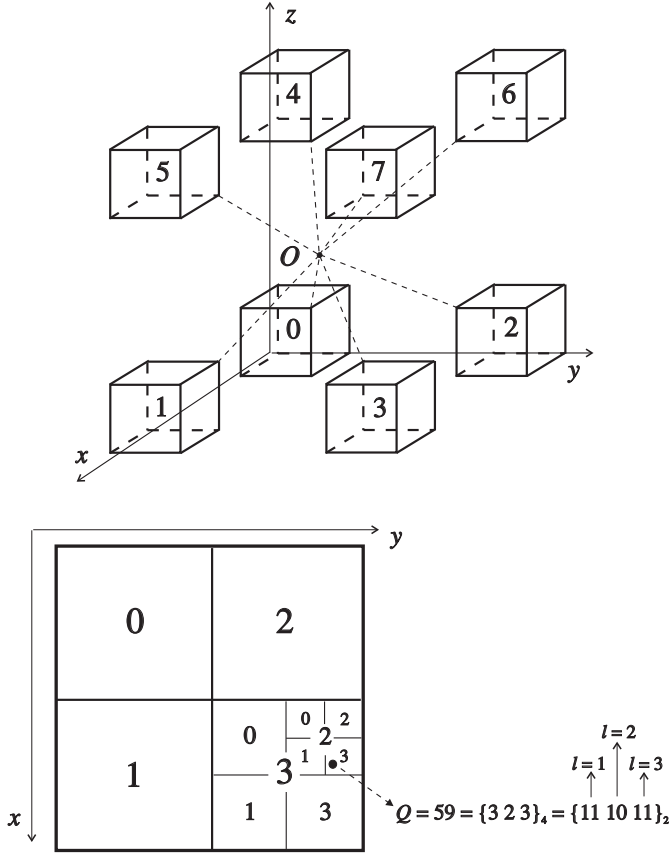}}
\caption{Top: a parent box (centered at $O$), has
been ``opened'' to show the numbering of its sub-boxes (in 3-D), which allows
to identify, by suitably extracting octal digits from the key $Q$ of a particle,
the sub-box which it belongs to. Bottom: example of
a particle mapping shown in the 2-D case for simplicity; in this case
($l_{\max}=3$) each sub-box enclosing the plotted particle, is recursively
identifies by base 4 digits made up of 2 bits.}
\label{sub-box}
\end{figure}

Given a particle's key, it is possible to determine quickly whether it belongs
to a box or not, in a recursive fashion. Let us indicate the key binary
representation as
$Q\equiv\{b_{3l_{\max}-1}b_{3l_{\max}-2}\cdot\cdot\cdot b_2b_1b_0\}_2$,
then: known that the particle belongs to a certain box at the level $l$ of the
spatial subdivision, at the level $l+1$ such particle will belong to the $i$-th
sub-box ($0\leq i \leq 7$) which is identified by the \emph{octal\/} digit
$i\equiv\{ b_{k+2}b_{k+1}b_k\}_2$. This latter is made up of the three
adjacent bits of $Q$ from position $k=3(l_{\max}-l-1)$ to the left.
For $l=0$ any particle is enclosed in the root box. See the example in
Fig.~\ref{sub-box}.
\noindent

Each key needs $3l_{\max}$ bits to be handled. We used two long-integers
(i.e. $16$ bytes), thus allowing $l_{\max}\leq 42$, which is sufficient for
most applications.
Note that the evaluation of the keys of all the particles, with a computational
cost of order $O(N)$, can be done just once, before the tree setting starts.

The mapping method described in WS is similar, to
some extent, to that above-illustrated, but in that case the authors do
not use pointers to reproduce the tree topology, they rather use the particles'
keys themselves (evaluated similarly as in Eq.\ref{Q}) plus a hashing function
to build an addressing space for all the boxes.
Roughly speaking, in order to reduce the huge number of a-priori possible keys
(e.g. in 3-D there are $8^{l_{\max}}$ possible values for $Q$), they truncate the
keys binary representation, replacing the information losted this way by means of
linked lists. In the authors opinion this presents mainly the following
advantages: i) it permits a direct access to any boxes, without needing a
tree-traversal; ii) in a distributed memory context, it gives a unique addressing scheme
for the boxes, independently of which PE owns them.
We think that the advantage in point i) is not important in our implementation
because, as we have seen, any procedure involving the tree structure is
performed \emph{recursively\/}. A direct access to a box does not
really speed up the execution; moreover, in WS' method the accesses are not really
direct (expecially at upper levels) due to the presence of linked lists. As far
as point ii) is concerned, also in our parallel version there is an addressing
scheme which allows data in any sub-domain to be globally referenced
(see Sect. \ref{par_tree_setting} and Appendix \ref{fulladdressing}).
\section{Construction of a global addressing scheme}
\label{fulladdressing}
Given the binary representation of
{\tt baddr}$\equiv\{b_{m-1} b_{m-2}\cdot\cdot\cdot b_1b_0\}_2$ that is the
address location of an upper box within the {\tt i}-th PE's local memory, we define
the box full-address as the $s$-bit number (being $s$ the bit size of integer variables,
excluding the bit used for the sign):
\begin{equation}
\mbox{\tt faddr}\equiv\{100\cdot \cdot \cdot 0b_{m-1}b_{m-2}\cdot\cdot\cdot b_1b_0
c_7c_6\cdot\cdot\cdot c_1c_0\}_2, \label{fulladdr}
\end{equation}
being
{\tt i}$\equiv\{c_7c_6\cdot\cdot\cdot c_1c_0\}_2$. Note that the leftmost bit,
in position $s-1$,
is set to 1 in order to indicate that the box pointed is an
\emph{upper\/} one. Note also that
8 bits are used for {\tt i}, allowing $p\le 256$. To make the full-address
be a single integer number it is necessary that $m+8<s$. Hence with 8-bytes
integers ($s=63$) the local addressing is limited by $2^{55}$ that is
sufficient for any purpose.

In FORTRAN such `bit concatenation' can be easily carried out this way:
\begin{equation}
\mbox{{\tt faddr = Ior(Ior(Ishft(baddr,8),i),mask)}},
\end{equation}
%
%
where {\tt mask}$=2^{s-1}$.
Inversely, given the full-address of a box, the following bit manipulation
instructions give the address in the local memory of the {\tt i}-th processor
which it belongs to:
\begin{eqnarray}
\mbox{\tt i} &=& \mbox{{\tt Iand(faddr,mask2)}},\label{fulladdr2}\\
\mbox{\tt baddr}&=& \mbox{{\tt Ishft(Iand(faddr,mask1),-8)}}
\label{fulladdr3}
\end{eqnarray}
being {\tt mask1}$=${\tt mask}$-1$ and {\tt mask2}$\equiv 255$.
\section{Parallel time integration}
\label{time_integration}
As far as the time integration is concerned, when the time advancing algorithm
adopts individual time steps, at a given time the forces are
evaluated only on a sub-set of all the particles. This leads to a
work-load as more unbalanced as the sub-set is smaller. We experimented various
possible solutions for such problem, in particular in the case of the \emph
{leapfrog} algorithm with the block-time scheme (Porter \cite{porter};
Henquist \& Katz \cite{HK}).
According to this scheme, the $i$-th particle occupies the `time bin'
$b_i\in\{0,1,2,...,b_{\rm max}\}$, in the sense that it gets a time step
$\Delta t_i=\tau/2^{b_i}$, where $\tau$ is the maximum time step allowed
(usually a fraction of a significant time scale for the system).
The simulation time ($t$) is advanced by the minimum time step used and, at a
given $t$, the accelerations are evaluated only on \emph{synchronized\/} particles, i.e.
on those whose acceleration was calculated last time at $t-\Delta t_i$
(at $t=0$ all the particles accelerations are evaluated).
According to some rules, $\Delta t_i$ can also change with time.

First, we found convenient to set the maximum a priori possible bin not that
high, say $b_{\rm max}<6$, so to avoid bins with too few particles within. This
choice is also recommendable because of the intrinsic non-symplectic nature of
the individual time step leapfrog. Indeed, every time a particle changes its bin,
a lost of time symmetry occurs, leading to instability and to a long term energy drift.
Furthermore, good results are obtained if, at a given instant, one assignes to
synchronized particles a weight ($w$) much greater than those assigned to the others.
One could be even tempted to set $w=0$ for non-synchronized particles because
no work will be done, in the current step, to update their accelerations. Nevertheless,
this would give rise to a very unbalanced work-load during the tree-setting, because
wide sets of PTERM boxes (containing zero-weight non-synchronized particles)
may be assigned to the same PE, forcing it to build large portions of the upper tree.
We found that a good compromise is to multiply the weight of the currently synchronized
particles  -- initially estimated as described in Sect. \ref{low-tree-setting} --  by the
factor $N/s$, being $s$ the number of these latter.

This mantains the unbalancing comparable with that of the single force evaluation on
all the particles, also in those highly dynamic situations
in which the accelerations of the particles moving through very dense region (like
for pairs of stars during close encounters occurring within the core of a
globular cluster) need to be frequently updated.

\end{document}